\documentclass[aps,preprint]{revtex4}
\usepackage{amsmath}
\usepackage{amssymb}
\usepackage{mathrsfs}
\usepackage{bm}
\usepackage{ulem}
\usepackage{xcolor}
\begin{document}
\preprint{CTP-SCU/2021010}	
\title{New Gedanken experiment on Reissner-Nordström AdS Black Holes surrounded by quintessence}

\author{Yang Qu}
\email{quyang@stu.scu.edu.cn}
\author{Jun Tao}
\email{taojun@scu.edu.cn}
\author{Jiayi Wu}
\email{wujiayi777@stu.scu.edu.cn}

\affiliation{Center for Theoretical Physics, College of Physics, Sichuan University, Chengdu, 610065, China}

\begin{abstract}
In this paper, we apply the new Gedanken experiment to investigate the weak cosmic censorship conjecture for Reissner-Nordström AdS black holes surrounded by quintessence. Since the perturbation of matter fields  doesn't affect the spacetime geometry, we propose the stability condition and assume the process of matter fields falling into the black hole satisfies the null energy condition.  Based on Iyer-Wald formalism we can derive the first order and second order variational idntities. From the two identities and the above two conditions lead to the first-order and second-order perturbation inequalities, and under the second-order approximation of matter fields perturbation, we find that the weak cosmic censorship conjecture is still satisfied.
\end{abstract}

\maketitle

\section{Introduction}
The singularity of the black hole is still a unsolve problem in general relativity. Black holes have the have singularity at the event horizon, but this is not a real singularity, the reason is that the coordinates we choose can not cover the whole spacetime manifold, we can choose suitable coordinates to remove this singularity. But inside the event horizon, there is a gravitational singularity, it will destroy the well-define spacetime and the law of causality. There is the weak cosmic censorship conjecture proposed by Penrose \cite{Penrose:1969pc} indecates that the conjecture indicates that the singularity  hidden inside the event horizon and one cannot detect the singularity. 

To prove this conjecture, the Gedanken experiment that was firstly proposed in \cite{wald1} by Wald to verify the weak cosmic censorship conjecture is vaild in extremal Kerr-Newman black holes. Over the last few years, many similar works based on this experiment have been considered in different works, for examples, different particles have been considered to verify the conjecture. In \cite{Cohen:1979zzb}, charged particles have been used to test the weak censorship conjecture in Reissner-Nordström (RN) black holes, a charged spining particle with parameters very much small have been used in \cite{Needham:1980fb}, in \cite{Semiz:1990fm}, the authors considered a spinless dyonic test particle, in \cite{Semiz:2005gs}, a weak packet of the complex massive scalar field replaced the particle to test the conjecture. More than this, the Gedanken experiment has been used in different black holes, a extreme RN black hole with a different type of $U(1)$ charge have been considered in \cite{Bekenstein:1994nx}, in \cite{Gwak:2017icn} the authors consider the warped AdS black hole based on 3-dimensional new massive gravity, moreover, the gauss-bonnet AdS black hole \cite{Zeng:2019hux} and the non-linear electrodynamics black hole \cite{Wang:2019jzz}. In addition, some other studies have also used this method \cite{Mu:2019bim,Isoyama:2011ea,Chen:2019pdj}. But the experiment only focus on the first-level perturbation of the process, if second-order or even higher-order perturbations are considered, the weak cosmic censorship conjecture may not be satisfied. Hubery \cite{Hubeny:1998ga} considered the second-order perturbation of test particles to show that the Kerr-Newman black holes can be destroyed in the Gedanken experiment. After that, the weak cosmic censorship conjecture for other kinds of black holes is examined, and it's found that the conjecture can be violated during the process \cite{deFelice:2001wj,Jacobson:2009kt,Rocha:2014jma,BouhmadiLopez:2010vc,Gao:2012ca,Duztas:2016xfg,Chirco:2010rq,Hod:2002pm,Saa:2011wq}.

To deal with these defects, Sorce and Wald \cite{Sorce:2017dst} proposed the new version of the Gedanken experiment, which considers matter fields instead of particles. Then an assumption indicates that a long enough time, the perturbation of the matter fields will not affect the background spacetime. Based on Iyer-Wald formalism\cite{Iyer:1994ys} and the null energy condition, the second-order perturbation inequality is derived to show that the weak cosmic censorship conjecture is valid. After that, many works based on this new experiment \cite{An:2017phb,He:2019mqy,Ge:2017vun,Chen:2019nhv,Jiang:2019ige,Jiang:2019vww,Jiang:2019soz,Wang:2020vpn,Jiang:2020mws,Wang:2019bml,Chen:2020hjm,Jiang:2020alh,Jiang:2020xow,Zhang:2020txy,Ding:2020zgg,Qu:2020nac} show that the conjecture is applicable for other kinds of black holes. In addition, there are other methods to study the weak cosmic censorship conjecture. In \cite{Bai:2020ieh,Gwak:2018akg,Gwak:2019asi,Gwak:2019rcz,Chen:2018yah,Chen:2019nsr}, the weak cosmic censorship conjecture was tested under the scattering of a scalar field. The authors consider two methods a massive scalar field and a test particle to study this conjecture in \cite{Yang:2020iat,Feng:2020tyc,Yang:2020czk,Eperon:2019viw}. The weak cosmic censorship conjecture was also studied by considering test particles or other different angles in \cite{Rocha:2011wp,Richartz:2008xm,Hod:2013oxa,Duztas:2013wua,Husain:2017cmj,Crisford:2017gsb,Yu:2018eqq,Gim:2018axz,Ghosh:2019dzq}.  

Observations over the last century have shown that the universe is dominated by an energetic component with  negative pressure \cite{Ostriker:1995su,Wang:1999fa}. One of the candidates for this component is  quintessence, which is a slow-changing and spatially non-uniform component of the negative pressure \cite{Ratra:1987rm,Frieman:1991tu,Chiba:1997ej,Turner:1998ex,Shaisultanov:1997bc,Bucher:1998mh}. The quintessence is described by an ordinary scalar field minimally coupled to gravity, whose potential leads to the expansion. Quintessence must be coupled to ordinary matter, which results in long-range interaction and time dependent natural constants, even if it's compressed to Planck's scale. Kiselev \cite{Kiselev:2002dx} proposed the typical dark energy with black holes to explain this phenomenon. Since then, relevant studies have been conducted successively \cite{Tsujikawa:2013fta,Ford:1987de,Fujii:1982ms}. We consider the contribution of the space-time gravitational field of extra energy, the momentum tensor, which consists of dark energy. Then we can study the thermodynamic effects of black holes surrounded by the dark energy \cite{Chen:2008ra,Fernando:2012ue,AzregAinou:2012hy,Chabab:2017xdw,Azreg-Ainou:2014lua,Hong:2019yiz,Liang:2020uul,Liang:2020hjz,Ghaffarnejad:2018bsd,Ghaffarnejad:2018exz,Ghaffarnejad:2020epjp,Yin:2021fsg}. Next, we will consider the thermodynamics of the black hole surrounded by quintessence and use the Gedanken experiment to test the the weak cosmic censorship conjecture.

Our paper is organized as follows. In Section \uppercase\expandafter{\romannumeral2}, we give a briefly introduction to RN-AdS black holes surrounded by quintessence. In Section \uppercase\expandafter{\romannumeral3}, we discuss the Iyer-Wald formalism and derive the  variational identities and the spacetime geometry of RN-AdS black holes under the perturbation of matter fields. In Section \uppercase\expandafter{\romannumeral4}, based on the variational identities and null energy condition, we derive the first-order and the second-order perturbation inequalities. In section \uppercase\expandafter{\romannumeral5} we prove that the weak cosmic censorship conjecture for nearly extremal RN-AdS black holes surrounded by quintessence is vaild under the second-order perturbation. In section \uppercase\expandafter{\romannumeral6}, some discussions and conclusions are given.

\section{RN-AdS black holes surrounded by quintessence}	
In this paper, we consider the RN-AdS black holes surrounded by quintessence in four-dimensional spacetime.  The action of this case could be described as follows 
\cite{Kiselev:2002dx,Ghaffarnejad:2018bsd}
\begin{equation}
\label{eq:1}
S=\frac{1}{16\pi}\int d^{4}x[(R-2\Lambda-F^{\mu\nu}F_{\mu\nu})+\mathcal{L}_{q}],
\end{equation}
where $F=dA$ is the strength of the electromagnetic field and $A$ is the potential of the electromagnetic field. $\Lambda$ is the cosmological constant with a negative value, $\mathcal{L}_{q}$ is the quintessence dark energy such as a barotropic perfect fluid and defined by \cite{Minazzoli:2012md}
\begin{equation}
\label{eq:2}
\mathcal{L}_{q}=-\rho_q\left[1+\omega \ln\frac{\rho}{\rho_0}\right],
\end{equation}
where $\rho_0$ is an integral constant, $\rho_q$ is the energy density, and $\omega$ is the quintessence dark energy barotropic index. The value range of $\omega$ is $-1<\omega<-\frac{1}{3}$ for the quintessence dark energy. The metric of RN-AdS black holes surrounded by quintessence is
\begin{equation}
\label{eq:3}
ds^{2}=-f(r)dt^{2}+\frac{1}{f(r)}dr^{2}+r^{2}(d\theta^{2}+sin^{2}\theta d\phi^{2}),
\end{equation}
where
\begin{equation}
\label{eq:5}
\begin{aligned}
f(r)=1-\frac{2M}{r}+\frac{Q^2}{r^2}-\frac{\Lambda r^2}{3}-\frac{a}{r^{3\omega+1}}.
\end{aligned}
\end{equation} 
 \textit{M} and \textit{Q}
are two integral constants which can be interpreted as the mass and charge of the black
hole respectively, \textit{l} is the radius of AdS space and is related to the cosmological constant by $\Lambda=-\frac{3}{l^2}$. The parameter \textit{a} is the normalization factor associated with the quintessence density of dark energy and it's always positive. The relationship between \textit{a} and $\rho_q$ can be expressed as \cite{AzregAinou:2012hy,Azreg-Ainou:2014lua}
\begin{equation}
\label{eq:6}
\rho_q=-\frac{a}{2}\frac{3\omega}{r^{3(\omega+1)}}.
\end{equation}
By imposing the condition $f(r)=0$, we can derive the radius of event horizon $r_+$. From the law of thermodynamics in the extended space \cite{AzregAinou:2012hy,Chabab:2017xdw,Hong:2019yiz}, we can obtain that
\begin{equation}
\label{eq:10}
P=-\frac{\Lambda}{8\pi},\quad V_+=\frac{4}{3}\pi r^3_+,\quad\mathcal{A}_+=-\frac{1}{2r_+^{3\omega}},
\end{equation} 
where \textit{P} is the thermodynamic pressure, \textit{V} is the thermodynamic volume, and $\mathcal{A}$ is conjugate to the parameter \textit{a}.

We consider the perturbative process of matter fields, the Lagrangian can be expressed as
\begin{equation}
\label{eq:8}
L=L_{EM}+L_{mt}.
\end{equation}
where $L_{EM}$ is Lagrangian of Einstein-Maxwell gravity, $L_{mt}$ is Lagrangian of matter field, and the spacetime geometry can be simply derived from the Lagrangian if the matter field satisfies the stable solution. 

The process of matter fields falling into the black hole can be considered as a completely dynamic process. We further assume that cosmological constant and normalization
factor \textit{a} are two effect parameters that relate to the matter source coupled to the Einstein-Maxwell gravity. Then we can obtain the energy-momentum tensor of matter fields \cite{Ghaffarnejad:2018bsd},
\begin{equation}
\label{eq:9}
T_{\mu\nu}=\frac{\Lambda}{8\pi}g_{\mu\nu}-\rho_qg_{\mu\nu}.
\end{equation}

In the next chapter, we will follow the basic works proposed by Wald \cite{Sorce:2017dst,Iyer:1994ys} to derive the black hole geometry under the perturbation of matter fields.
\section{The linear variational identities}
In this paper, we would like to use the Iyer-Wald formalism \cite{Sorce:2017dst,Iyer:1994ys} to investigate the Gedanken experiments in RN-AdS black holes surrounded by quintessence. Based on the Iyer-Wald formalism \cite{Sorce:2017dst}, $n$-form Lagrangian can be regarded as the function of metric $g_{\mu\nu}$ and other tensor fields $\Psi$ in spacetime. The dynamical field can be represented by $\varphi$, i.e. $\varphi=(g_{\mu\nu},\Psi)$. In this case, we set $\varphi$ to be the collection of dynamical field $\varphi=(g_{\mu\nu},A)$ and use $\lambda$ to express the perturbation of the matter field to $\varphi$.  Consider the Einstein-Maxwell Theory, 
the variation of Lagrangian is
\begin{equation}
\label{eq:17}
\delta L=E_{\varphi}\delta \varphi +d\vartheta(\varphi,\delta \varphi),
\end{equation}
where $E_{\varphi}=0$ gives the equations of motion. $\vartheta$ is the three-form of the symplectic potential and is locally composed of $\varphi$ and its derivatives. In the Einstein-Maxwell Theory, $\vartheta$ can be linearly expressed with gravitational field part and electromagnetic field part as
\begin{equation}
\label{eq:18}
\begin{aligned}
&\Theta^{GR}_{\mu\nu\sigma}(\varphi,\delta \varphi)=\frac{1}{16\pi}\varepsilon_{\rho\mu\nu\sigma}g^{\rho \gamma}g^{\alpha\beta}(\nabla_{\beta}\delta g_{\gamma\alpha}-\nabla_{\gamma}\delta g_{\alpha\beta}),\\&\Theta^{EM}_{\mu\nu\sigma}(\varphi,\delta \varphi)=-\frac{1}{4\pi}\varepsilon_{\rho\mu\nu\sigma}F^{\rho\gamma}\delta A_{\gamma}.
\end{aligned}
\end{equation}
Then we define the three-form of symplectic current as
\begin{equation}
\label{eq:19}
w(\varphi,\delta_1 \varphi,\delta_2 \varphi)=\delta_1\vartheta(\varphi,\delta_2 \varphi)-\delta_2\vartheta(\varphi,\delta_1 \varphi).
\end{equation}
It also can be linearly expressed by two parts,
\begin{equation}
\label{eq:20}
\begin{aligned}
&w^{GR}_{\mu\nu\sigma}=\frac{1}{16\pi}\varepsilon_{\rho\mu\nu\sigma}\eta^{\rho},\\&w^{EM}_{\mu\nu\sigma}=\frac{1}{4\pi}(\delta_2 (\varepsilon_{\rho\mu\nu\sigma}F^{\rho\gamma})\delta_1 A_{\gamma}-\delta_1(\varepsilon_{\rho\mu\nu\sigma}F^{\rho\gamma})\delta_2 A_{\gamma}),
\end{aligned}
\end{equation}
where
\begin{equation} 
\label{eq:21}
\eta^\mu=\mathcal{P}^{\mu\nu\sigma\rho\gamma\alpha}(\delta_2g_{\nu\sigma}\nabla_{\rho}\delta_1g_{\gamma\alpha}-\delta_1g_{\nu\sigma}\nabla_{\rho}\delta_2g_{\gamma\alpha}),
\end{equation}
with
\begin{equation}
\label{eq:22}
\mathcal{P}^{\mu\nu\sigma\rho\gamma\alpha}=g^{\mu\gamma}g^{\alpha\nu}g^{\sigma\rho}-\frac{1}{2}g^{\mu\rho}g^{\nu\gamma}g^{\alpha\sigma}-\frac{1}{2}g^{\mu\nu}g^{\sigma\rho}g^{\gamma\alpha}-\frac{1}{2}g^{\nu\sigma}g^{\mu\gamma}g^{\alpha\rho}+\frac{1}{2}g^{\nu\sigma}g^{\mu\rho}g^{\gamma\alpha}.
\end{equation}
For an arbitrary vector field, the Noether current associated with $\zeta^a$ is defined by
\begin{equation}
\label{eq:23}
J_{\zeta}=\vartheta(\varphi,\mathscr{L}_\zeta \varphi)-\zeta \cdot L.
\end{equation}
The variation of Noether current \cite{Iyer:1994ys}  is 
\begin{equation}
\label{eq:24}
\delta J_{\zeta}=-\zeta \cdot (E(\varphi) \cdot\delta\varphi)+ w(\varphi,\delta \varphi,\mathscr{L}_\zeta \varphi)+d(\Theta(\zeta \cdot\varphi,\delta \varphi)).
\end{equation}
As shown in Ref. \cite{Iyer:1995kg}, we can write the Noether current as
\begin{equation}
\label{eq:25}
J_{\zeta}=C_{\zeta}+dQ_{\zeta},
\end{equation} 
where $Q_{\zeta}$ is the Noether charge and $C_{\zeta}=\zeta^{\mu}C_{\mu}$ is the constraint to the theory. We can obtain $ C_{\mu}=0$ and $dJ_{\zeta}=0$ from the equation of motion. The Noether charge $Q_{\zeta}$ is linearly expressed by
\begin{equation}
\label{eq:26} 
Q_{\zeta}=Q^{GR}_{\zeta}+Q^{EM}_{\zeta},
\end{equation}
where
\begin{equation}
\label{eq:27}
\begin{aligned}
&(Q^{GR}_{\zeta})_{\mu\nu}=-\frac{1}{16\pi}\varepsilon_{\mu\nu\sigma\rho}\nabla^{\sigma}\zeta^{\rho},\\&(Q^{EM}_{\zeta})_{\mu\nu}=-\frac{1}{8\pi}\varepsilon_{\mu\nu\sigma\rho}F^{\sigma\rho}A_{\gamma}\zeta^{\gamma}.
\end{aligned}
\end{equation}
Considering the Einstein-Maxwell Theory, the equations of motion and constraints are given by
\begin{equation}
\label{eq:28}
\begin{aligned}
&E_{\phi}\delta \phi=-\varepsilon(\frac{1}{2}T^{\mu\nu}\delta g_{\mu\nu}+j^{\mu}\delta A_{\mu}),\\&C_{\mu\nu\sigma\rho}=\varepsilon_{\gamma\nu\sigma\rho}(T^{\gamma}_{\mu}+A_{\mu}j^{\gamma}),
\end{aligned}
\end{equation}
where $T_{\mu\nu}=\frac{1}{8\pi}G_{\mu\nu}-T^{EM}_{\mu\nu}$ and $j^b=\frac{1}{4\pi}\nabla_{\mu}F^{\nu\mu}$ are the energy-momentum tensor and electric current respectively.

By differentiating Eq.(\ref{eq:25}) and substituting Eq.(\ref{eq:24}), the first-order variational identity can be expressed by
\begin{equation}
\label{eq:29}
d(\delta Q_{\zeta}-\zeta \cdot \Theta(\varphi,\delta \varphi))=w(\varphi,\delta \varphi,\mathscr{L}_\zeta \varphi)-\zeta \cdot E_\varphi \delta \varphi-\delta C_\zeta.
\end{equation}
In the same way, by differentiating Eq.(\ref{eq:29}), the second-order variational identity can be shown as
\begin{equation}
\label{eq:30}
d(\delta^2 Q_{\zeta}-\zeta \cdot \delta \Theta(\varphi,\delta \varphi))=w(\varphi,\delta \varphi,\mathscr{L}_\zeta \delta \varphi)-\zeta \cdot\delta E_\varphi \delta \varphi-\delta^2 C_\zeta.
\end{equation}

Considering the perturbation process of matter fields, for RN-AdS black holes surrounded by quintessence, the spacetime field equations are
\begin{equation}
\label{eq:11}
\begin{aligned}
&G_{\mu\nu}(\lambda)=8\pi(T^{EM}_{\mu\nu}(\lambda)+T_{\mu\nu}(\lambda)),\\&
\bigtriangledown_{\mu}^{(\lambda)}F^{\nu\mu}(\lambda)=4\pi j^{\nu}(\lambda),
\end{aligned}
\end{equation}
where $T^{EM}_{\mu\nu}$is the energy-momentum tensor of electromagnetic field, $T_{\mu\nu}$ is the energy-momentum tensor of matter fields.

We will use the Eddingdon-Finkelstein coordinate, compared with Schwarzschild gauge, there is no singularity at the event horizon, therefore when we consider the matter fields following into the black hole, it is convenient to use the gauge to calculate. we arrive at the spacetime geometry of the following \cite{Wang:2019bml} 
\begin{equation}
\label{eq:12}
ds^2=-f(r,v,\lambda)dv^2+2\mu(r,v,\lambda)drdv+r^2(d\theta^2+sin^2\theta d\phi^2).
\end{equation}

We assume that the spacetime satisfies the stability condition\cite{Sorce:2017dst}, which means that after enough long time, the matter fields perturbation does not affect thr spacetime geometry, the metric should be consistent with the RN-AdS black hole surrounded by quintessence. We utilize $M$, $Q$, $\lambda$, $a$ to describe the properties of black holes and represent these quantities by the parameter $\lambda$. Thus, the dynamical fields can be expressed as \cite{Wang:2019bml}
\begin{equation}
\label{eq:13}
\begin{aligned}
&ds^2=-f(r,\lambda)dv^2+2drdv+r^2(d\theta^2+sin^2\theta d\phi^2),\\&
F(\lambda)=dA(\lambda),
\end{aligned}
\end{equation}
where
\begin{equation}
\label{eq:14}
f(r,\lambda)=1-\frac{2M(\lambda)}{r}+\frac{Q^2(\lambda)}{r^2}-\frac{\lambda (\lambda)r^2}{3}-\frac{a(\lambda)}{r^{3\omega+1}},\quad A(\lambda)=-\frac{Q(\lambda)}{r}dv
\end{equation}
The energy-momentum tensor of matter fields is 
\begin{equation}
\label{eq:15}
T_{\mu\nu}(\lambda)=\left[\frac{\Lambda(\lambda)}{8\pi}-\rho_q(\lambda)\right]g_{\mu\nu}(\lambda).
\end{equation}

When the parameter $\lambda$ is zero, the spacetime should still be the solution of RN-AdS black holes surrounded by quintessence, i.e. when $f(r,0)=f(r)$, metric can be restored to the original spacetime form.
\section{FIRST-ORDER AND SECOND-ORDER PERTURBATION INEQUALITIES}
In this section, we calculate the integral of first-order and second-order variational identities to obtain the perturbation inequalities. Due to the stability condition, we can choose a hypersurface $\Sigma=H \cup \Sigma_1$. \textit{H} is a portion of the horizon $r=r_+$ in the background spacetime, starting from the unperturbed horizon's bifurcate surface \textit{B} and ending up on the cross-section $B_1$. $\Sigma_1$ approaches infinity along the time-slice ($\nu$=constant) as a space-like hypersurface. And the event horizon is a Killing horizon generated by the Killing field $\zeta^a$. For the first-order variational equation, we utilize the condition $\mathscr{L}_\zeta \varphi=0$ and integrate it on the hypersurface $\Sigma$
\begin{equation}
\label{eq:31}
\int_{\Sigma}d(\delta Q_{\zeta}-\zeta \cdot \Theta(\varphi,\delta \varphi))+\int_{\Sigma}\zeta \cdot E_\varphi \delta \varphi+\int_{\Sigma}\delta C_\zeta=0.
\end{equation}

Using the Stokes theorem and the condition that \textit{B} is unperturbed horizon's bifurcate surface, it can be rewritten as 
\begin{equation}
\label{eq:32}
\int_{S_o}(\delta Q_{\zeta}-\zeta \cdot \Theta(\varphi,\delta \varphi))+\int_{\Sigma_1}\zeta \cdot E_\varphi \delta \varphi+\int_{\Sigma_1}\delta C_\zeta+\int_{H}\delta C_\zeta=0.
\end{equation}
Since the integral diverges as the integral region approaches infinity, we apply a cut-off method at sphere $S_o$ with radius $r_o$ and let the limit of $S_o$ approach asymptotic infinity.

Then, we calculate each integration term separately. Firstly, we evaluate the first term of Eq.(\ref{eq:32}). Considering Eq.(\ref{eq:13}), Eq.(\ref{eq:18}) and Eq.(\ref{eq:27}),  we have
\begin{equation}
\label{eq:36}
\begin{aligned}
\int_{S_{o}}(\delta Q-\zeta\cdot\Theta(\varphi,\delta\varphi))&=\int_{S_{o}}(\delta Q^{GR}-\zeta\cdot\Theta^{GR}(\varphi,\delta\varphi))+\int_{S_{o}}(\delta Q^{EM}-\zeta\cdot\Theta^{EM}(\varphi,\delta\varphi))\\&=\delta M-V_{o}\delta P-\mathcal{A}_{o}\delta a-\frac{1}{8\pi}\int_{S_{o}}\epsilon_{\mu\nu\sigma\rho}A_{\gamma}\zeta^{\gamma}\delta F^{\sigma\rho}.
\end{aligned}
\end{equation}
As the $S_e$ approaching asymptotic infinity, the second term vanishes, and then the first term of Eq.(\ref{eq:32}) is given by
\begin{equation}
\label{eq:41}
\int_{S_o}(\delta Q_{\zeta}-\zeta \cdot \Theta(\varphi,\delta \varphi))=\delta M-V_o\delta P-\mathcal{A}_o\delta a.
\end{equation}

Secondly, we integrate the second part of Eq.(\ref{eq:32}) by considering $\varphi(0)$ as a globally hyperbolic, asymptotically flat solution of the equations of motion, which means that the integral of the second term equal to zero \cite{Sorce:2017dst}.

Then, we utilize the condition $j^a=0$ on $\Sigma_1$ and Eq.(\ref{eq:28}) to show that
\begin{equation}
\label{eq:34}
\begin{aligned}
\int_{\Sigma_1}\delta C_\zeta=(V_o-V_+)\delta P+(\mathcal{A}_o-\mathcal{A}_+)\delta a,
\end{aligned}
\end{equation}
where
\begin{equation}
\label{eq:35}
\begin{aligned}
&V_e=\frac{4}{3}\pi r_o^3,\quad\mathcal{A}_o=-\frac{1}{2r_o^{3\omega}}.&
\end{aligned}
\end{equation}
Finally, with $A_a\zeta^a|_H=-\Phi_+$ and $\int_{H}\varepsilon_{ebcd} \delta j =\delta Q$, we can further obtain the expression as
\begin{equation}
\label{eq:42}
\int_{H}\delta C_\zeta=\int_{H}\varepsilon_{\gamma\nu\sigma\rho}\zeta^a{\mu}delta T^{\gamma}_{\mu}-\Phi_+ \delta Q.
\end{equation}

Since both the normal vectors $n^a$ and the time-like Killing vectors $\zeta^{\mu}$ on the horizon become null and $\zeta^{\mu}\propto n^{\mu}$, we can use the null energy condition $\delta T_{\mu\nu}n^{\mu}n^{\nu}\geqslant 0$. On the horizon, we have $\varepsilon_{\gamma\nu\sigma\rho}=-4n_{[\gamma}\widetilde{\varepsilon}_{\nu\sigma\rho]}$, where $n^\mu$ is the normal vector and $\widetilde{\varepsilon}_{\nu\sigma\rho}$ is the volume element, so the first term of Eq.(\ref{eq:42}) can be written as $\int_{H}\widetilde{\varepsilon}_{\nu\sigma\rho}n_{\gamma}\zeta^{\mu}\delta T_{\mu} ^{\gamma}(\lambda) \geqslant 0$. Therefore, from Eq.(\ref{eq:41},\ref{eq:34},\ref{eq:42}) and the null energy condition, we can obtain the first-order perturbation inequality
\begin{equation}
\label{eq:43}
\delta M-\Phi_+ \delta Q-V_+\delta P-\mathcal{A}_+\delta a \geqslant 0.
\end{equation}

In our work, we use the second-order perturbation inequality to examine whether the weak cosmic censorship conjecture can be violated under the second-order perturbation approximation of matter fields. The weak cosmic censorship conjecture is valid under the first-order approximation when the first-order inequality is satisfied,. But when the first-order perturbation takes the optimal option 
\begin{equation}
\label{eq:44}
\delta M-\Phi_+\delta Q-V_+\delta P-\mathcal{A}_+\delta a  =0,
\end{equation}
the conjecture can not be examined sufficiently by only considering the first-order approximation. Then we need to derive the second-order perturbation inequality.

Integrating Eq.(\ref{eq:30}) on the hypersurface $\Sigma$, it yields
\begin{equation}
\label{eq:45}
\begin{aligned}
&\int_{S_o}\delta(\delta Q_{\zeta}-\zeta \cdot \Theta(\varphi,\delta \varphi))+\int_{\Sigma}\delta(\zeta \cdot E_\varphi \delta \varphi)+\int_{\Sigma_1}\delta^2 C_\zeta+\int_{H}\delta^2 C_\zeta-\mathscr{W}_{H}(\varphi,\delta \varphi)-\mathscr{W}_{\Sigma_1}(\varphi,\delta \varphi)=0,\\&
\end{aligned}
\end{equation}
where
\begin{equation}
\label{eq:46}
\begin{aligned}
&\mathscr{W}_{H}(\varphi,\delta \varphi)=\int_{H}w(\varphi,\delta \varphi,\mathscr{L}_\zeta \delta \varphi),\\&\mathscr{W}_{\Sigma_1}(\varphi,\delta \varphi)=\int_{\Sigma_1}w(\varphi,\delta \varphi,\mathscr{L}_\zeta \delta \varphi).
\end{aligned}
\end{equation}
Following the previous calculation steps, the second term of Eq.(\ref{eq:45}) equals to 0, the third and fourth term of Eq.(\ref{eq:45}) can be expressed respectively
\begin{equation}
\begin{aligned}
&\label{eq:47}
\int_{\Sigma_1}\delta^2 C_\zeta=(V_e-V_+)\delta^2 P+(\mathcal{A}_e-\mathcal{A}_+)\delta^2 a,\\&\int_{H}\delta^2 C_\zeta=\int_{H}\varepsilon_{\gamma\nu\sigma\rho}\zeta^{\mu}\delta^2 T_{\mu} ^{\gamma}(\lambda)-\Phi_+ \delta^2 Q.
\end{aligned}
\end{equation} 
Similarly, the first term can be written as
\begin{equation}
\label{eq:48}
\begin{aligned}
\int_{S_{o}}\delta(\delta Q-\zeta\cdot\Theta(\varphi,\delta\varphi))&=\int_{S_{o}}\delta(\delta Q^{GR}-\zeta\cdot\Theta^{GR}(\varphi,\delta\varphi))+\int_{S_{o}}\delta(\delta Q^{EM}-\zeta\cdot\Theta^{EM}(\varphi,\delta\varphi))\\&=\int_{S_{o}}\delta(\delta Q^{GR}-\zeta\cdot\Theta^{GR}(\varphi,\delta\varphi))-\frac{1}{8\pi}\int_{S_{o}}\epsilon_{\mu\nu\sigma\rho}\delta A_{\gamma}\zeta^{\gamma}\delta F^{\sigma\rho}\\&-\frac{1}{8\pi}\int_{S_{o}}\epsilon_{\mu\nu\sigma\rho}A_{\gamma}\zeta^{\gamma}\delta^{2}F^{\sigma\rho}\\&=\delta^{2}M-V_{o}\delta^{2}P-A_{o}\delta^{2}a-\frac{1}{8\pi}\int_{S_{o}}\epsilon_{\mu\nu\sigma\rho}\delta A_{\gamma}\zeta^{\gamma}\delta F^{\sigma\rho},
\end{aligned}
\end{equation} 
where the second term on the last line vanishes if $S_{e}$ approaches asymptotic infinity, and then the first term of Eq.(\ref{eq:45}) is given by
\begin{equation}
\label{eq:49}
\int_{S_o}\delta Q_{\zeta}-\zeta \cdot \Theta(\varphi,\delta \varphi)=\delta^2 M-V_o\delta^2 P-\mathcal{A}_o\delta^2 a.
\end{equation}
Finally, we calculate the fifth term of Eq.(\ref{eq:45}). It can be linearly expressed by two parts
\begin{equation}
\label{eq:51}
\mathscr{W}_{H}=\int_{H} w^{GR}+\int_{H} w^{EM}.
\end{equation}
From \cite{Sorce:2017dst}, we can obtain the similar result, and then we have
\begin{equation}
\label{eq:53}
\mathscr{W}_{H}=\int_{H}\varepsilon_{\gamma\nu\sigma\rho}\zeta^{\mu}\delta^2T_{\mu}^{\gamma EM}.
\end{equation}

Therefore, from Eqs. (\ref{eq:47}),  (\ref{eq:48}), and (\ref{eq:53}), we can get
\begin{equation}
\label{eq:54}
\delta^2M-\Phi_+ \delta^2 Q-V_+\delta^2 P-\mathcal{A}_+\delta^2 a=\mathscr{W}_{\Sigma_1}(\varphi,\delta \varphi)-\int_{H}\varepsilon_{\gamma\nu\sigma\rho}\zeta^{\mu}\delta^2(T_{\mu}^{\gamma EM}+T_{\mu}^{\gamma}).
\end{equation}

Following the same method in Ref. \cite{Sorce:2017dst}, we can build an auxiliary spacetime to calculate. Because of the stability condition, the spacetime geometry on $\Sigma_1$ is still the RN-AdS spacetime surrounded by quintessence, and the configuration of dynamical fields under the perturbation of matter field can be described by one parameter $\lambda$, and we only consider the first-order variation of matter field on auxiliary spacetime,  then the $M^{RA}(\lambda)$ , $Q^{RA}(\lambda)$ , $\Lambda^{RA}(\lambda)$ and $a^{QR}(\lambda)$ are given as follows,
\begin{equation}
\begin{aligned}
\label{eq:57}
M^{QR}(\lambda)&=M+\lambda \delta M, \ \ Q^{QR}(\lambda)=Q+\lambda \delta Q, 
\\ \Lambda^{QR}(\lambda)&=\Lambda+\lambda \delta \Lambda,\ \ \  a^{QR}(\lambda)=a+\lambda\delta a.
\end{aligned}
\end{equation}
Considering only the first-order variation of the matter field, we obtain $\delta^2M^{QR}=\delta^2Q^{QR}=\delta^2\Lambda^{QR}=\delta^2a^{QR}=0$. Then we obtain $\delta \varphi^{QR}=\delta \varphi$ on hypersurface $\Sigma_1$,  which implies that  $\mathscr{W}_{\Sigma_1}(\varphi,\delta \varphi)=\mathscr{W}_{\Sigma_1}(\varphi,\delta \varphi^{QR})$.  Thus we can calculate them straightly in auxiliary spacetime.

Integrating the second-order variation identity on $\Sigma_1$, we have
\begin{equation}
\label{eq:58}
\begin{aligned}
\mathscr{W}_{\Sigma_1}(\varphi,\delta \varphi^{QR})=\int_{\partial \Sigma_1}\delta(\delta Q^{QR}_{\zeta}-\zeta \cdot \Theta(\varphi^{QR},\delta \varphi^{QR})).
\end{aligned}
\end{equation}

Following the previous calculation steps and the gauge condition of the electromagnetic field such that $\zeta^a\delta A_a=0$ at \textit{H}, the Eq.(\ref{eq:58}) can be expressed by
\begin{equation}
\label{eq:59}
\mathscr{W}_{\Sigma_1}(\varphi,\delta \varphi^{QR})=\frac{1}{4\pi}\int_{B_1}\frac{\delta Q^2}{r}sin\theta d\theta d\phi=\frac{\delta Q^2}{r_+}.
\end{equation}

Then, the Eq.(\ref{eq:54}) can be rewritten as
\begin{equation}
\label{eq:60}
\delta^2M-\Phi_+ \delta^2 Q-V_+\delta^2 P-\frac{\delta Q^2}{r_+}-\mathcal{A}_+\delta^2a=-\int_{H}\varepsilon_{\gamma\nu\sigma\rho}\zeta^{\mu}\delta^2(T_{\mu}^{\gamma EM}+T_{\mu}^{\gamma}).
\end{equation}

We consider the null energy condition  $\delta^2 (T^{EM}_{\mu\nu}+T_{\mu\nu})n^{\mu}n^{\nu}\geqslant 0$. under the second-order approximation. Then the second-order perturbation inequality can be reduced as
\begin{equation}
\label{eq:61}
\delta^2M-\Phi_+ \delta^2 Q-V_+\delta^2 P-\frac{\delta Q^2}{r_+}-\mathcal{A}_+\delta^2a\geqslant 0.
\end{equation}

\section{Test the weak cosmic censorship conjecture of RN-AdS black hole surrounded by quintessence}
In this section, we will apply the new version of the Gedanken experiment to discuss the weak cosmic censorship conjecture of nearly extremal RN-AdS black holes surrounded by quintessence. We assume that the spacetime satisfies the stability condition. The condition of the existing event horizon $r_+$ is metric factor satisfy $f(r_+)=0$. We suppose that there exists one minimum point at $r = r_0$ for $f(r)$, and the existence of the event horizon is consistent with condition $f(r_0)\leqslant0$. We can take $f(r_0(\lambda),\lambda)$ and use the discriminant function to represent the change of extremum of $f(r)$ under the matter field perturbation. $r_0(\chi)$ is the minimum point of $f(r_0(\lambda),\lambda)$, which satisfies the condition $\partial_r f(r_0(\lambda),\lambda)=0$. We can expand the function to second-order at $\lambda=0$,
\begin{equation}
\label{eq:62}
f(r_0(\lambda),\lambda)\backsimeq f(r_0,0)+ f'\lambda+ f''\frac{\lambda^2}{2}+\mathcal{O}(\lambda^2).
\end{equation} With $\partial_r f(r_0(\lambda),\lambda)=0$ and the zero-order approximation of $\lambda$, one can obtain
\begin{equation}
\label{eq:63}
M = \frac{6Q^2r_0^{3\omega-1}+2\Lambda r_0^{3\omega+3}-3(3\omega+1)a}{6r_0^{3\omega}}.
\end{equation}
Considering the matter fields and taking the first-order variation of  $\partial_r f(r_0(\lambda),\lambda)=0$, we have
\begin{equation}
\label{eq:64}
\delta r_0 = \frac{2r_0^{3\omega+1}}{2\Lambda r_0^{3\omega+3}+3\omega(3\omega+1)a-2Q^2r_0^{3\omega-1}}\left[\delta M - \frac{2Q\delta Q}{r_0}-\frac{\delta \Lambda r_0^3}{3}+\frac{(3\omega+1)\delta a}{2r_0^{3\omega}}\right].
\end{equation}

Therefore, by applying Eqs.(\ref{eq:63}) and (\ref{eq:64}), we can yield the detailed expression of Eq.(\ref{eq:62}) as
\begin{equation}
\label{eq:65}
\begin{aligned}
f(r_0(\lambda),\lambda)&=\frac{r_0^{3\omega+1}-Q^2r_0^{3\omega-1}-\Lambda r_0^{3\omega+3}+3\omega a}{r_0^{3\omega+1}}\\&-\frac{2\lambda}{r_0}\left(\delta M-\frac{Q\delta Q}{r_0}+\frac{\delta \Lambda r_0^3}{6}+\frac{\delta a}{2r_0^{3\omega}}\right)\\&-\frac{\lambda^2}{r_0}\left\lbrace \delta^2 M-\frac{Q\delta^2 Q}{r_0}+\frac{\delta^2 \Lambda r^3_0}{6}+\frac{\delta^2 a}{2r_0^{3\omega}}+\left[r_0\Lambda+\frac{3\omega(3\omega+1)a}{2r_0^{3\omega+2}}-\frac{Q^2}{r_0^3}\right]\delta r_0^2\right\rbrace \\&+\lambda^2\left[\frac{\delta Q^2+2\delta M\delta r_0}{r_0^2}-\frac{4Q\delta Q\delta r_0}{r_0^3}-\frac{2r_0\delta \Lambda \delta r_0}{3}+\frac{(3\omega+1)\delta a\delta r_0}{r_0^{3\omega+2}}\right].
\end{aligned}
\end{equation}

The event horizon $r_+$ and $r_0$ satisfy the relation $r_+(1-\varepsilon)=r_0$ with $\varepsilon \ll 1$ \cite{Sorce:2017dst} for the case of the nearly extremal black hole. With the relation of $\partial_rf(r_0)=0$, $f'(r_+)=\varepsilon r_+f''(r_+)$, and we can obtain the relation of $f(r_0)=-\frac{1}{2}\varepsilon^2 r_+^2f''(r_+)$ under the second-order approximation of $\varepsilon$. We can conclude these relations,

\begin{equation}
\label{eq:66}
\frac{r_0^{3\omega+1}-Q^2r_0^{3\omega-1}-\Lambda r_0^{3\omega+3}+3\omega a}{r_0^{3\omega+1}}=\left[1-\frac{2Q^2}{r_+^2}+\frac{3\omega(3\omega+3)a}{2r_+^{3\omega+1}}\right]\varepsilon^2.
\end{equation}

Therefore, we can rewrite the expression of Eq.(\ref{eq:65}) as
\begin{equation}
\label{eq:fr67}
\begin{aligned}
f(r_0(\lambda),\lambda)&=\left[1-\frac{2Q^2}{r_+^2}+\frac{3\omega(3\omega+3)a}{2r_+^{3\omega+1}}\right]\varepsilon^2\\&-\frac{\lambda \varepsilon}{r_+^2}(\delta M-\Phi_+\delta Q-V_+\delta P-\mathcal{A}_+\delta a)+\frac{\lambda \varepsilon}{r_+^2}\left(2Q\delta Q+r_+^4\delta \Lambda-\frac{3\omega \delta a}{r_+^{3\omega-1}}\right)\\&-\frac{\lambda^2}{r_0}\left\lbrace (\delta^2M-\Phi_+ \delta^2 Q-V_+\delta^2 P-\frac{\delta Q^2}{r_+}-\mathcal{A}_+\delta^2a)+\left[r_0\Lambda+\frac{3\omega(3\omega+1)a}{2r_0^{3\omega+2}}-\frac{Q^2}{r_0^3}\right]\delta r_0^2\right\rbrace \\&+\lambda^2\left[\frac{\delta Q^2+2\delta M\delta r_0}{r_0^2}-\frac{4Q\delta Q\delta r_0}{r_0^3}-\frac{2r_0\delta \Lambda \delta r_0}{3}+\frac{(3\omega+1)\delta a\delta r_0}{r_0^{3\omega+2}}\right].
\end{aligned}
\end{equation}
Utilizing Eq.(\ref{eq:44}) and Eq.(\ref{eq:61}) together with above results, we can further simplify the expression as
\begin{equation}
\label{eq:67}
\begin{aligned}
f(r_0(\lambda),\lambda)&\leq\left[1-\frac{2Q^2}{r_+^2}+\frac{3\omega(3\omega+3)a}{2r_+^{3\omega+1}}\right]\varepsilon^2+\frac{\lambda \varepsilon}{r_+^2}\left(2Q\delta Q+r_+^4\delta \Lambda-\frac{3\omega \delta a}{r_+^{3\omega-1}}\right)\\
&+\frac{\lambda^2(-2Q\delta Qr_+^{3\omega-1}-\delta \Lambda r_+^{3\omega+3}+3\omega\delta a)^2}{2r_+^{3\omega+1}(2\Lambda r_+^{3\omega+3}+3\omega(3\omega+1)a-2Q^2r_+^{3\omega-1})}.
\end{aligned}
\end{equation}
With the condition that $f((1+\varepsilon)r_0)=0$ and $f'(r_0)=0$, and considering the zero-order approximation of $\varepsilon$, we can derive $\Lambda$ and M,
\begin{equation}
\label{eq:68}
\begin{aligned}
&\Lambda=\frac{r_0^{3\omega+1}-Q^2r_0^{3\omega-1}+3\omega a}{r_0^{3\omega+3}},\\&M = \frac{2r_0^{3\omega+1}+4Q^2r_0^{3\omega-1}-3(\omega+1)a}{6r_0^{3\omega}}.
\end{aligned}
\end{equation}
Together with the relation $r_+=(1+\varepsilon)r_0$, we can expressed Eq.(\ref{eq:67}) as
\begin{equation}
\label{eq:69}
\begin{aligned}
f(r_0(\lambda),\lambda) \leq -\frac{r_0^3(\varepsilon A+\lambda B)^2}{2f''(r_0)},
\end{aligned}
\end{equation} 
where
\begin{equation}
\label{eq:70}
\begin{aligned}
&A=9 a \omega ^2+9 a \omega -4 Q^2 r_0^{3 \omega -1}+2 r_0^{3 \omega +1},\\
&B=2Q\delta Qr_0^{3\omega-1}+\delta \Lambda r_0^{3\omega+3}-3\omega\delta a.
\end{aligned}
\end{equation}

Because $r_0$ is the minimum point that satisfies the condition $f''(r_0)>0$. The above expression gives $f(r_0(\lambda),\lambda)\leqslant 0$, which means that after a second-order perturbation, the event horizon for the quintessence surrounding the nearly extreme RN-AdS black hole still exists, thus the weak cosmic censorship conjecture is unable to be violated.

\section{Conclusion}
In this paper, we discuss the weak cosmic censorship conjecture of nearly extremal RN-AdS black holes surrounded by quintessence with the new version of the Gedanken experiment. Most forms of energy, due to gravity, will cause the universe to expansion to slow down ,but for quintessence, it can accelerate the expansion of the universe. We consider RN-AdS black holes in the quintessence as barotropic perfect fluid, the variable related to the quintesssence can be regarded as thermodynamic variables in the extended space. We can obtian the spacetime geometry and the corresponding thermodynamics about RN-AdS black holes surrounded by quintessence.

Consider the matter fields perturbation process, we assume that matter fields will not affect the space-time geometry after a long time perturbation, then we propose the stability assumption and  think the peocess of matter fields falling into the black hole should satisfy the null energy condition, then utilize Iyer-Wald formalism to derive the first-order and the second-order perturbation inequalities. Under the condition of meeting the first-order optimal option, we use the second-order inequality to prove that the event horizon of a nearly extreme RN-AdS black hole surrounded by quintessence does not vanish, which implies that the conjecture can not be violated for this case. This result is the same as that of previous papers\cite{Hong:2019yiz,Liang:2020uul,Liang:2020hjz}. Furthermore, we can prove whether the event horizon still exists under the higher perturbation or another black hole, this gives us a broader perspective and methods to examine the weak cosmic censorship conjecture.

\begin{acknowledgments}
We are grateful to Peng Wang, Wei Hong, and Siyuan Hui for useful discussions. This work is supported in part by NSFC (Grant No.11947408 and 11875196 and 12047573).
\end{acknowledgments}

\end{document}